# Phase-field Simulations of Polarization Variations in Polycrystalline $Hf_{0.5}Zr_{0.5}O_2$ based MFIM: Voltage-Dependence and Dynamics

Revanth Koduru, Imtiaz Ahmed, Atanu K. Saha, Xiao Lyu, Peide Ye, and Sumeet K. Gupta
Purdue University, West Lafayette, IN, USA

*Abstract*— In this work, we investigate the device-to-device variations in remanent polarization of Hafnium-Zirconium-Oxide based Metal-Ferroelectric-Insulator-Metal (MFIM) stacks. We consider the effects of polycrystallinity in conjunction with multi-domain effects in HZO to understand the dependencies of variations on static and dynamic voltage stimuli using our 3D dynamic multi-grain phase-field simulation framework. We examine the trends in variations due to various design factors – set voltage, pulse amplitude and pulse width and correlate them to the dynamics of polarization switching and the underlying mechanisms. According to our analysis, variations exhibit a non-monotonic dependence on $V_{SET}$ due to the interplay between voltage-dependent switching mechanisms and the polycrystalline structure. We further report that towards the higher end of the set voltages, collapsing of oppositely polarized domains can lead to increase in variations. We also show that ferroelectric thickness scaling lowers the device-to-device variations. In addition, considering the dynamics of polarization switching, we signify the key role of voltage and temporal dependence of domain nucleation in dictating the trends in variations. Finally, we show that to reach a target mean polarization, using a pulse with lower amplitude for longer duration results in lower variations compared to a higher amplitude pulse of a shorter duration.

## I. INTRODUCTION

Ferroelectric devices have gained renewed attention in recent years due to the discovery of ferroelectricity in doped Hafnium Oxide [1] and its CMOS process compatibility [2], thickness scalability [3] and other appealing attributes [4-9]. Researchers have demonstrated multiple flavors of Hafnium-Zirconium-Oxide (HZO)-based ferroelectric (FE) devices, including ferroelectric capacitors, ferroelectric field effect transistors (FEFETs), and ferroelectric tunnel junctions (FTJs). HZO-based devices have been shown as promising candidates for various emerging and in-demand applications such as non-volatile memories, electronic synapses, and neurons [9-14]. However, commercialization of these devices is contingent upon addressing several key material and device-level challenges [15-20]. Among these, the device-to-device variations [20-25] resulting from the polycrystalline nature [26] of HZO is particularly crucial due to its significant implications on a wide range of applications [20-23]. Certainly, optimizing the material properties of HZO via process and material engineering will be key in addressing the polycrystallinity and stability of multiple phases [20]. Nonetheless, understanding the dependence of the device-to-device variations on various design knobs and device parameters is crucial for developing superior design solutions.

  Previous studies in this area [21-25] have investigated various dependencies of polycrystallinity-induced device-to-device variations and proposed various material and design-level optimization strategies. For example, Ni et al. [21] showed an increase in variations due to FEFET area scaling. They also showed that higher pulse amplitudes and longer pulse widths lower the variations and improve the distinguishability between negative and positive ($+P$ and $-P$) polarization states. Koduru et al. [25] showed that the variations decrease with decreasing FE layer thickness and correlated this trend with the increase in domain density and reduction in random domain nucleation processed with FE thickness scaling. Ni et al. [24] examined the impact of variations in ferroelectric layer and transistor parameters in FEFETs and found that FE parameter variations contribute significantly to the device variations.

  However, most of these variation studies have only considered the two-state ($+P$ and $-P$) operation of the FE layer, while multi-state [12, 13, 33] ($+P$, $-P$ and intermediate polarization states) operation of FE devices is crucial for various applications such as multi-state memories, synapses, and neurons. In the context of multi-state operation of FEFETs, experiments [23, 31] have shown a non-monotonic dependence of variations in the polarization states on the set or reset voltages. The underlying mechanisms responsible for this non-monotonic dependence are not yet fully understood. Also, a comprehensive correlation between variations and the effects of underlying polycrystalline structure and polarization switching mechanisms is still lacking. Further, the dependence of variations on voltage pulse parameters, such as pulse amplitude and width and their combined effect has not yet been fully explored for multi-state operation.

In this work, we aim to address these gaps by extensively analyzing the impact of applied voltage pulse on polycrystallinity-induced device-to-device variations. To achieve this, we utilize our in-house 3D dynamic multi-grain phase-field simulation framework. This framework captures the multi-domain polarization switching dynamics in HZO along with the effects of polycrystallinity as well as inter- and intra-grain interactions. We consider polycrystalline HZO-based Metal-Ferroelectric-Insulator-Metal (MFIM) stacks for our analysis, as it is a primitive structure across various FE devices. We focus on analyzing the variations in the remanent polarization of samples due to its crucial role in determining the distinguishability between different states in FE-based devices. Key contributions of this work include:

- A thorough analysis of the dependence of device-to-device variations on the set voltage ($V_{SET}$) considering multi-state polarization retention.
- Investigation of the dependence of the device-to-device variations on voltage pulse parameters, specifically pulse amplitude ($V_{pulse}$) and pulse width ($t_{pulse}$).
- Presentation of a strategy to achieve a target mean intermediate polarization state across samples with minimal device-to-device variations.
- Correlation of trends in variations to the multi-domain polarization switching mechanisms and the effects of underlying polycrystalline structures.

## II. 3D DYNAMIC MULTI-GRAIN PHASE-FIELD SIMULATION FRAMEWORK

Our 3D multi-grain phase-field simulation framework [25] (Fig. 1(a)) captures the effects of polycrystalline nature of HZO [26] along with the multi-domain formation [33] on the polarization characteristics of MFIM stacks (Fig. 1(b)). We achieve this by coupling a 3D grain-growth equation [27] that models the polycrystalline structure of HZO, with a 3D phase-field model that simulates the behavior of the MFIM stack subjected to electric field or voltage ($V_{APP}$).

The 3D grain-growth equation [27] models the polycrystalline microstructure using multiple abstract order parameters ($\eta_k$) whose evolution follows a time-dependent Ginzburg Landau formalism. We then randomly map these parameters ($\eta_k$) to the grain orientation angles ($\theta_i$) – the angle between the polarization direction ($c$-axis in the orthorhombic crystal phase [34]) of the grain and the physical thickness direction of the HZO film (Fig. 1(d)). The grain orientation angle ($\theta_i$) serves as the differentiating factor between the grains (Fig. 1(c)), resulting in different polarization directions across the grains.

The 3D phase-field model [25] calculates the potential ($\phi$) and polarization ($P$) profiles of the MFIM stack by self-consistently solving the Time-dependent Ginzburg Landau (TDGL) and Poisson's equations for a given applied voltage ($V_{APP}$). The TDGL equation models the polarization switching behavior of the FE layer, taking into account the free, gradient, and depolarization energies, as well as the surface energy at the FE-DE interface. Poisson's equation captures the electrostatic behavior of the system, accounting for the polarization-induced bound charges in the FE layer. We self-consistently solve these equations in 3D real space using the finite-difference method and considering the polycrystalline structure generated by the grain-growth equations for the ferroelectric layer.

Including the effects of polycrystallinity in TDGL and Poisson's equations necessitates accounting for the variations in the polarization direction across the grains. In our framework, we accomplish this by using two coordinate systems: a global coordinate system ($x, y, z$) and a local coordinate system ($a, b, c$) as shown in Fig. 1(d). The $z$-axis of the global coordinate system aligns with the physical thickness direction of the MFIM stack. On the other hand, the $c$-axis of the local coordinate system aligns with the polarization direction of the grain, making angle $\theta_i$ with the z-axis. Thus, the local coordinate system varies from one grain to another, capturing the varying polarization direction. We use the global coordinate system to solve Poisson's equation in the entire stack and the local coordinate system to solve the TDGL equation in the FE layer. The variables in the local and global coordinate systems are related via the rotation matrix ($R$) described in Fig. 1(e). Thus, our framework captures the multi-domain polarization switching in conjunction with the multi-grain attributes of HZO. By virtue of the TDGL equations, polarization switching via both domain nucleation and domain growth are captured. Further, electrostatic interactions between different grains is also accounted for in our model.

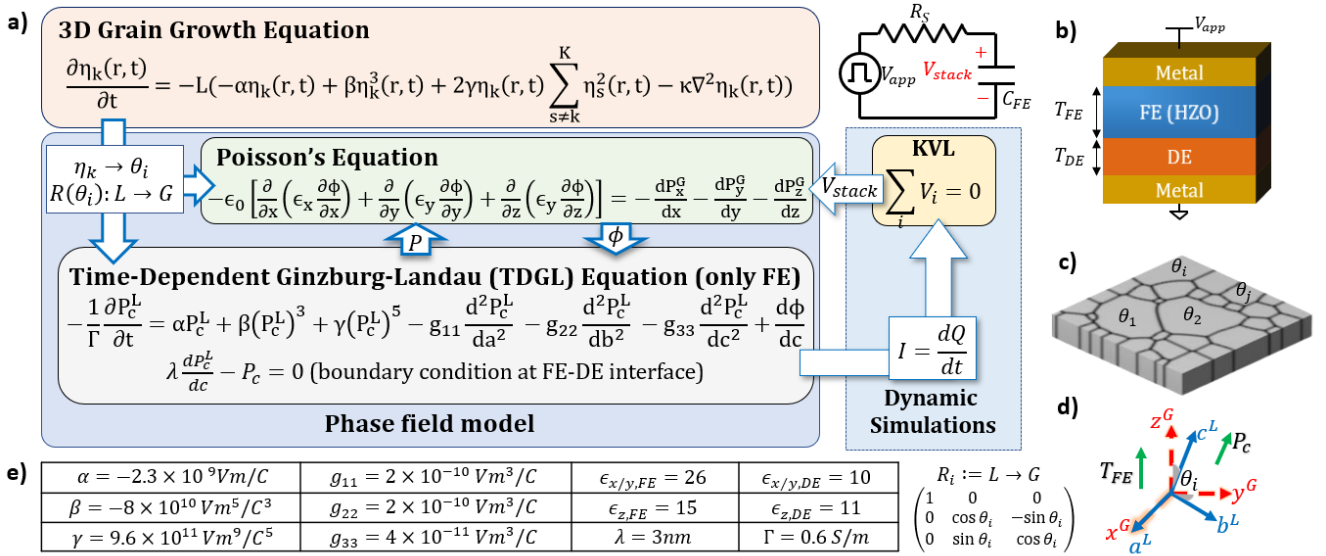

Fig. 1. a) Dynamic 3D grain-growth coupled phase-field simulation framework (Poisson's +TDGL self consistently coupled with KVL) considering the equivalent circuit emulating the measurement setup. b) Metal-Ferroelectric-Insulator-Metal structure. c) Sample polycrystalline structure generated by the 3D grain-growth equation. d) Two coordinate axes system used in the framework – $x$-$y$-$z$ (global) and $a$-$b$-$c$ (local). e) Calibrated parameters used in the simulation framework.

Further advancing our simulation framework, we incorporated the dynamics of polarization switching by coupling the phase-field model with a "Kirchhoff's Voltage Law (KVL)" module. The KVL module is implemented considering a Thevenin's equivalent for the circuit used in experimental setup (Fig. 2(c)). For a given applied voltage ($V_{APP}$) profile, the KVL module obtains the current ($I = dQ/dt$) flowing in the circuit at each time step and calculates the voltage across the MFIM stack ($V_{stack}$). We then pass this voltage ($V_{stack}$) to the phase-field model, which determines the polarization ($P$) and potential ($\phi$) profiles of the MFIM stack. To ensure consistency in the temporal evolutions of polarization ($P$), current ($I$), and potential ($\phi$), we solve the phase-field and KVL models in an iterative and self-consistent manner at each time step.

Our framework is distinctive in its ability to capture non-uniform grain shapes and sizes along with inter- and intra- grain interactions, unlike most other models for HZO-based devices. Further, it also captures polarization switching through both domain growth and domain nucleation mechanisms. However, it is important to note that, in our current framework, we make certain assumptions such as the presence of only the orthorhombic phase in the FE layer, similar inter- and intra-grain elastic interactions and uniform strain in the HZO layer. Additionally, we do not consider the effects of traps and defects. While these assumptions will change the magnitude of variations, we expect that the trends in variations that we present (which is the main objective of this paper) and their correlations to the underlying physical mechanisms will hold.

To ensure the accuracy in the trends predicted by our framework, we calibrate and validate the grain-growth and phase-field models using experimental data. We validate the grain growth equations by comparing the grain diameter distributions of the generated polycrystalline samples to the experimental distributions for various HZO film thicknesses, as reported in [25].

For the phase-field model, we fabricate MFIM samples with 3nm $Al_2O_3$ interfacial layer and 5 and 10nm HZO ferroelectric layers. These samples are fabricated using atomic layer deposition (ALD) at $200^0C$ followed by rapid thermal annealing at $500^0C$ in an $N_2$ environment (details of fabrication methodology in [29]). We then characterize these samples using low-frequency Q-V measurements to calibrate the material permittivity ($\epsilon$), free energy ($\alpha, \beta, \gamma$), and gradient energy ($g_{11}, g_{22}, g_{33}$) coefficients of the TDGL equation. To factor in the device-to-device and cycle-to-cycle variations in the Q-V characteristics, we simulate 20 MFIM samples with different polycrystalline structures for the HZO layer over five voltage cycles and use the average of these simulated characteristics to calibrate with the experimental data. The parameters are calibrated for major loop of 10nm $T_{FE}$. Using the *same* parameters, we simulate the minor loops for $T_{FE}$=10nm as well as other MFIM structures with $T_{FE}$=7 and 5nm. As shown in Fig. 2(a, b, c), the simulated and experimental characteristics show a good agreement, which validates the trends across different $T_{FE}$ and applied voltages.

We also calibrate the viscosity coefficient ($\Gamma$) of the TDGL equation by measuring the temporal evolution of polarization using the Positive-Up, Negative-Down (PUND) technique. Specifically, we use the experimental setup and pulse scheme described in [30] and reproduced in Fig. 2(c, d). The resulting evolution of simulated normalized polarization ($P/P_s$) shows a good match with experimental data [30] (Fig. 2(e)). The calibrated TDGL and material parameters are summarized in Fig. 1(e).

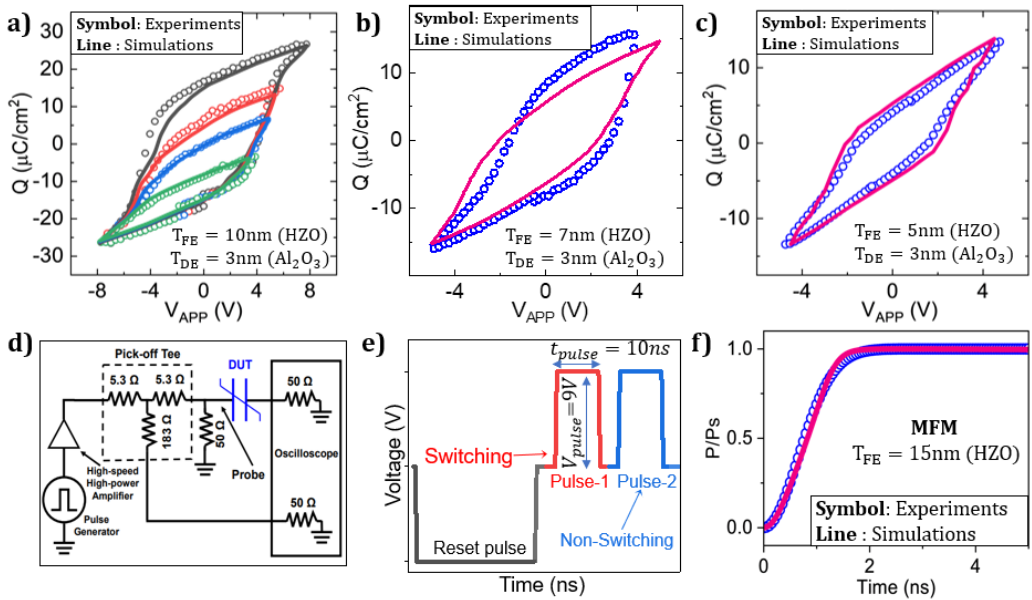

Fig. 2. Simulated and experimental charge ($Q$) versus applied voltage ($V_{APP}$) characteristics for MFIM stack for two different $T_{FE}$ of a) 10nm, b) 7nm, and c) 5nm. d) Experimental setup [30] for dynamic measurements using PUND technique and e) Pulse scheme used for dynamic measurements in experiments [30]. f) Simulated and experimental normalized polarization ($P/P_s$) versus time ($t$).

### III. DEVICE-TO-DEVICE VARIATIONS

In this section, we investigate the impact of applied voltage ($V_{APP}$) on the polycrystallinity-induced device-to-device variations in the MFIM stack. We further correlate the variations to the underlying polarization switching mechanisms and the multi-domain polarization profiles. Additionally, we analyze the effect of ferroelectric thickness ($T_{FE}$) on variations. To conduct this study, we employ a two-stage approach.

*Stage-1:* We first use quasi-static simulations to study the variations by sweeping the applied voltage ($V_{APP}$) across the MFIM stack in small incremental steps and allowing samples to reach a steady state at each step. This stage enables us to understand the effect of the maximum applied voltage (referred to as the set voltage ($V_{SET}$)) on the variations decoupled from the time-dependent effects.

*Stage-2:* We then use dynamic simulations to examine the impact of sharp voltage changes and the limited duration of voltage application, which is more representative of the typical use cases. Through this approach, we study the dependence of variations on the voltage pulse parameters i.e., pulse amplitude ($V_{pulse}$) and pulse width ($t_{pulse}$).

In both the stages of our study, we simulate 100 MFIM samples with a 3nm $Al_2O_3$ interfacial layer and a 10nm HZO ferroelectric layer unless otherwise specified. These samples are composed of different polycrystalline structures for the ferroelectric layer capturing variations in the number, size, shape and orientation of the grains. To qualitatively understand the dependencies of device-to-device variations, we focus on presenting the distributions and the relevant statistical parameters of remanent polarization ($P_R$) since $P_R$ plays a central role in the distinguishability of memory states in FEFETs, FERAMs and FTJs. It is important to note that in addition to device-to-device variations, a single sample may exhibit variations in the device characteristics across multiple voltage cycles [25]. This is known as cycle-to-cycle variations or stochasticity and is primarily due to the stochastic nature of the domain nucleation. To minimize the effect of cycle-to-cycle variations in our analysis of device-to-device variations, the trends and results presented are averaged over five voltage cycles.

#### A. Quasi-Static analysis

We begin by analyzing the impact of set voltage ($V_{SET}$) on variations by quasi-statically simulating the MFIM samples under a slowly varying $V_{APP}$ (as in Fig. 3(a)). The samples are simulated over a range of $V_{SET}$ values while maintaining a fixed reset voltage ($V_{RESET}$) of -8V. We specifically focus on the variations in the positive remanent polarization ($P_{R+}$) i.e., polarization at 0V during the backward path as $V_{APP}$ decreases from $V_{SET}$.

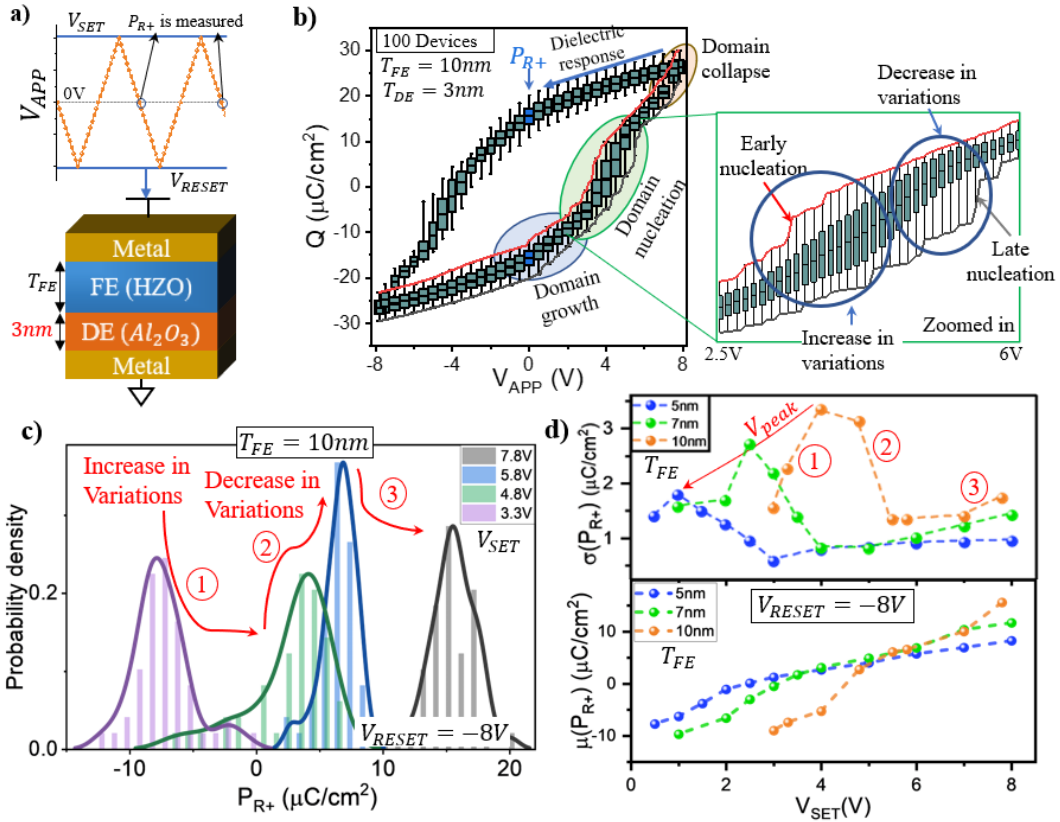

Fig. 3. a) MFIM stack structure and the applied voltage ($V_{APP}$) pulse scheme used for quasi-static simulations. b) $Q$-$V_{APP}$ characteristics depicting the device-to-device variations at each voltage step (via box plot) with dominant polarization switching mechanisms highlighted, zoomed in Q-V in the domain-nucleation region highlighting the early and late nucleation across samples. c) Statistical distribution of variations in $P_{R+}$ for different set voltages ($V_{SET}$) showing the non-monotonic dependence of variations. d) Standard deviation ($\sigma(P_{R+})$) and mean ($\mu(P_{R+})$) of variations in $P_{R+}$ versus $V_{SET}$ showing non monotonic trend in variations and peak point voltage ($V_{peak}$) for different FE thickness ($T_{FE}$).

The statistical distributions of $P_{R+}$ for different $V_{SET}$ values (Fig. 3(c)) reveal a non-monotonic trend in the device-to-device variations as a function of $V_{SET}$. The standard deviation of these variations ($\sigma(P_{R+})$) versus $V_{SET}$ shown in Fig. 3(d) highlights this trend. As $V_{SET}$ increases, an initial increase in the variations (labeled as region-1) is observed. This is followed by a decrease in $\sigma(P_{R+})$ (region-2) after $V_{SET}$ surpasses a critical voltage ($V_{peak}$). As $V_{SET}$ continues to increase, we see a slight increase in the variations again (region-3). It is worth pointing out that similar non-monotonic dependence of device-to-device variations on $V_{SET}$ has been observed in experiments with HZO-based ferroelectric devices [23, 31], albeit in a different structure.

Before we gain insights into the non-monotonic trends, let us briefly review the polarization-switching mechanisms in HZO. The polarization-switching in HZO is primarily driven by two mechanisms: domain growth (domain-wall motion) and domain nucleation. When $V_{APP}$ increases, domain growth-based switching leads to an increase in the size of the existing $+P$ polarized domains. This results in a gradual change in polarization and smooth $Q$-$V_{APP}$ characteristics (see domain growth in Fig. 3(b)). In contrast, domain-nucleation involves the formation of new $+P$ polarized domains when the electric field in a region exceeds a critical value. This leads to an abrupt change in polarization and steep $Q$-$V_{APP}$ characteristics (see domain-nucleation in Fig. 3(b)). Notably, domain nucleation is a random process, dependent on the underlying polycrystalline structure as well as the multi-domain polarization and potential profiles.

To understand the mechanisms behind the non-monotonic trend, we will analyze the evolution of the underlying polarization domain profiles at the $P_{R+}$ state as $V_{SET}$ increases. Fig. 4 shows the $P_{R+}$ domain profiles as $V_{SET}$ increases for representative two samples (sample-1 and sample-2). Due to variations in their underlying polycrystalline structures, the coercive voltage of sample-1 ($V_{C1}$) is lower than that of sample-2 ($V_{C2}$). Generalizing this observation to multiple samples, variations in the underlying polycrystalline structures lead to a non-uniform distribution of coercive voltage, with a significant proportion of the samples clustered around the mean [25].

As $V_{SET}$ increases from 0V, the initial change in polarization is dominated by domain growth (until 3.3V in Fig. 4), as the electric field in the FE layer is below the critical field required for nucleation. However, as $V_{SET}$ continues to increase, the polarization-switching in a small subset of samples transitions to domain nucleation (sample-1 between 3.3V and 4.8V) as the electric field in these samples surpasses the critical threshold. This transition depends on various factors such as underlying polycrystalline structure and polarization domain profiles. This results in randomness in the polarization switching across the samples. The dominance of nucleation in this subset of samples leads to a sharp change in their polarization, while the majority of samples continue a gradual polarization change due to domain growth (sample-2 between 3.3V and 4.8V). In other words, the increasing gap in polarization between the nucleating and non-nucleating samples start to widen the distribution of $P_{R+}$ (increase in variations in Fig. 3(b)) and increase the device-to-device variations, as in region-1.

As $V_{SET}$ continues to increase, two opposing effects come into picture. The larger number of domains in the early nucleating samples results in rapid change of polarization in these samples compared to the domain growth dominated ones. This tends to further increase the device-to-device variations. On the other hand, an increasing number of samples starts to nucleate with the increasing $V_{SET}$. These samples try to catch up the early nucleating ones striving to reduce the variations. But as long as the nucleation-driven samples are a minority compared to domain growth-dominated samples, the device-to-device variations continue to increase (as in region-1 Fig. 3(c, d)) with $V_{SET}$. Therefore, this initial increase in variations with $V_{SET}$ is mainly due to random domain nucleation and in turn, the random but sharp polarization change induced by it, which amplifies the variations in the underlying polycrystalline structures across the samples.

As $V_{SET}$ increases beyond a critical voltage ($V_{peak}$), majority of samples transition to domain nucleation (e.g. sample-2 after 4.8V in Fig. 4). Additionally, the early nucleating samples have already switched a significant amount of polarization i.e. large areas of these samples switched to +P domains (sample-1 after 4.8V). This reduces the scope of further domain nucleation in these samples and causes them to transition back to domain growth. Domain growth leads to a gradual polarization change in these samples, allowing the late nucleating samples to catch up. As a result, the distribution of polarization starts to narrow (labeled in Fig. 3(b)), reducing the device-to-device variations (as in region-2).

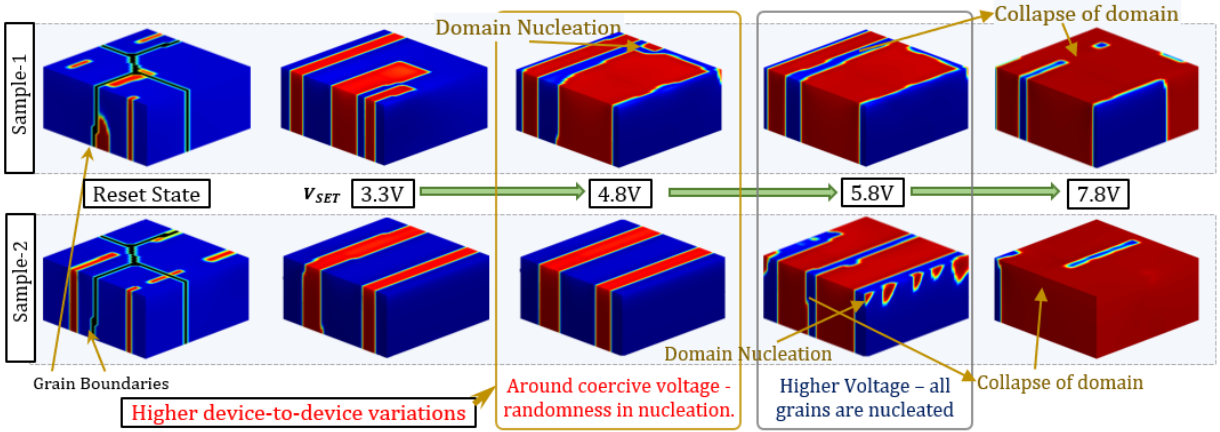

Fig. 4: Evolution of polarization domain profiles at the $P_{R+}$ state (with the underlying polycrystalline structure embedded in the left most figure) starting from reset state with the increasing set voltage contrasting the underlying polarization switching mechanisms (domain growth and domain nucleation) and variations in two different samples.

With further increase in $V_{SET}$, domain growth results in the gradual expansion of +P domains, while the oppositely polarized -P domains shrink reaching a critical size. As a result, even a small increase in the applied voltage collapses these -P domains – abruptly switching them to +P regions. The point at which the domains collapse depends on the underlying polycrystalline structures and polarization domain profiles. Hence, this introduces a level of randomness in the switching behavior across the samples. Additionally, the collapse of domains switches a marginally higher polarization compared to pure domain growth. These two factors contribute to the slight increase in device-to-device variations as observed in the region-3.

Interestingly, we observe that the non-monotonic trend in the variations with $V_{SET}$ remains consistent across different FE thicknesses ($T_{FE}$) – 5, 7, and 10nm ($\sigma(P_{R+})$ vs $V_{SET}$ in Fig. 3(d)). However, as $T_{FE}$ decreases, we see a decrease in the maximum variations and a shift in the critical voltage ($V_{peak}$) to lower values. The reduction in peak variations can be attributed to the evolution of denser domain patterns with $T_{FE}$ scaling, resulting from the interplay between gradient and depolarization energies (details in [33]). These denser domain patterns reduce the scope of domain nucleation, and domain growth becomes the dominant polarization-switching mechanism. Domain growth leads to a gradual polarization-switching, which in-turn reduces the amplification of underlying polycrystalline variations. This results in the decrease of peak device-to-device variations with $T_{FE}$ scaling. Further, the electric field in the FE layer surpasses the critical field required for nucleation at lower $V_{SET}$ for scaled $T_{FE}$ [33]. This results in the nucleation of new domains at lower $V_{SET}$ and, in turn, the reduction of $V_{peak}$ with $T_{FE}$ scaling.

Although not explicitly shown here, we observe that reset voltage ($V_{RESET}$) has a similar impact on the device-to-device variations as the set voltage ($V_{SET}$), but with regard to the negative remanent polarization $P_{R-}$, which is associated with the reset state. This is attributed to the symmetry of Q-V loops. Specifically, we observe a non-monotonic trend in the variations of $P_{R-}$, as $V_{RESET}$ decreases from 0V. This trend is characterized by an initial increase in variations due to early nucleation of -P domains in a small subset of samples, which amplifies the underlying polycrystalline variations. As $V_{RESET}$ decreases and crosses the critical voltage, the variations decrease as most samples transition to domain nucleation and the early nucleating samples shift to domain growth allowing the late nucleating ones to catch up. As $V_{RESET}$ continues to decrease, the trend concludes with a slight increase in variations due to the random collapse of +P polarization domains.

## B. Dynamic Analysis

Next, we investigate the impact of voltage pulses on the device-to-device variations by dynamically simulating the MFIM samples with varying pulse amplitudes and widths (Fig. 5(a)). To begin, we reset the samples by applying a reset pulse of -8V for a prolonged duration – allowing the samples to reach a steady state. We then apply a sharp switching pulse with varying pulse widths ($t_{pulse}$) ranging from 2ns to 100ns and amplitudes ($V_{pulse}$) ranging from 4V to 10V. After removing the switching pulse, we analyze the variations in the polarization state of the samples, which we refer to as the remanent polarization ($P_{R+}$). Specifically, we investigate the impact of $V_{pulse}$ and $t_{pulse}$ on device-to-device variations in $P_{R+}$.

The results of our simulations reveal a significant dependence of variations in remanent polarization ($\sigma(P_{R+})$) on pulse width ($t_{pulse}$) and amplitude ($V_{pulse}$), as in Fig. 5(b). Specifically, we observe that, for all pulse amplitudes, the variations are high initially and decrease with increasing pulse width (labeled as region-1 in Fig. 5(b))). However, for pulse amplitudes ≤6V, variations start to increase after certain pulse width (region-2a in Fig. 5(b)). Whereas for higher amplitudes (of 8V-10V), variations continue to decrease as the pulse width increases (region-2 b in Fig. 5(b)). To better understand these trends, let us look at the underlying polarization domain profiles and correlate the variations to the effects of polycrystallinity and polarization switching mechanisms.

We begin by analyzing the initial decrease of variations in $P_R$ with increasing pulse width (region-1 in Fig. 5(b)). The initially high variations observed at shorter $t_{pulse}$ are primarily due to the polycrystallinity-induced variations in the switching times across the samples. To understand this, consider two grains with different orientation angles ($\theta_i$). When subjected to voltage along the z-axis, the grain with a lower orientation angle experiences a stronger electric field ($E_Z \cos(\theta_i)$) in the polarization direction. As a result, in accordance with Merz's law [35], this grain with lower orientation angle responds faster than the one with the higher orientation angle.

When considering multiple samples, due to the variations in underlying polycrystalline structures, some samples respond faster to the applied voltage pulse than others. As a result, for the shorter pulse widths, these faster responding samples switch a large amount of polarization than the slower ones, leading to higher device-to-device variations. Additionally, the variations are observed to be higher for higher pulse amplitudes ($V_{pulse}$) as the higher $V_{pulse}$ leads to dominance of domain nucleation, due to the electric field in the FE layer crossing the critical threshold for nucleation (Fig. 5 (d) 10V at 3ns). This leads to the switching of a relatively large amount of polarization in the samples than lower $V_{pulse}$, which amplifies the underlying polycrystalline variations resulting in relatively high device-to-device variations for higher $V_{pulse}$.

As the pulse width increases, the slower samples get adequate time to respond to the applied pulse. Additionally, the rate of polarization change is typically more rapid at the onset of switching and gradually decreases as the switching progresses ($\mu(P_{R+})$ vs. $t_{pulse}$ in Fig. 5(c)). As a result, the polarization switching in the faster-responding samples slows down, allowing the slower samples to catch up and reducing the overall device-to-device variations. Furthermore, the decrease in variations with increasing pulse width is more pronounced for higher $V_{pulse}$ due to the initial high variations. Moreover, higher $V_{pulse}$ yields shorter switching times, which lead to a quicker response from the samples to the applied voltage. Therefore, at higher $V_{pulse}$ the slower responding samples catch up to the faster ones earlier compared to lower $V_{pulse}$. This results in the pronounced reduction of the device-to-device variations with $t_{pulse}$ for higher pulse amplitudes(region-1 in Fig. 5(b)).

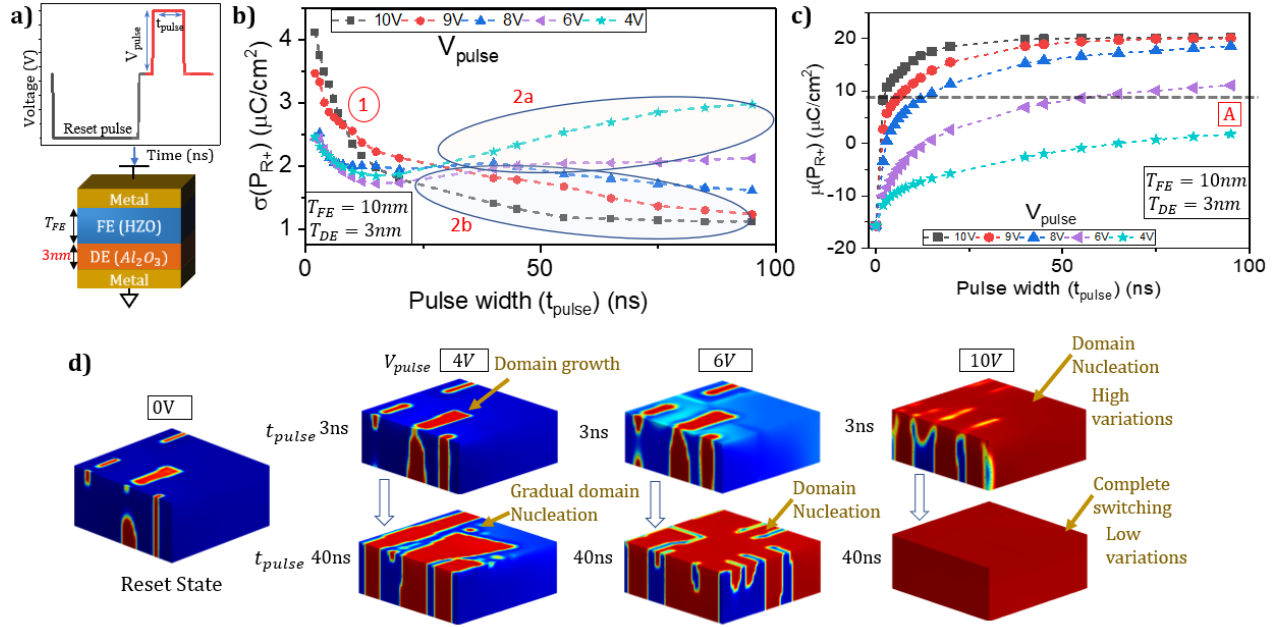

Fig. 5. a) MFIM stack with the voltage pulse scheme used. b) Standard deviation ($\sigma(P_{R+})$) and c) mean ($\mu(P_{R+})$) of device-to-device variations versus pulse width ($t_{pulse}$) for different pulse amplitudes ($V_{pulse}$). d) Polarization domain structures at reset state and $V_{pulse}$ of 4,6 and 10V for $t_{pulse}$ of 3ns and 40ns showing the differences in the underlying polarization switching mechanisms.

As the pulse width increases further, we observe that the variations continue to decrease for higher $V_{pulse}$ (10, 9 and 8V in Fig. 5(b) region-2b). This is because, at such high pulse amplitudes for long durations, the samples start to switch completely and reach a single domain state (Fig. 5(d) 10V at 40ns). This results in the standard deviation of variations ($\sigma(P_R)$) saturating and being purely due to the polycrystallinity-induced variations in the grain shape and orientations (i.e., with minimal effect of multi-domain dynamics).

On the other hand, for lower pulse amplitudes ($\leq$ 6V in Fig. 5(b) region-2a), we see that the variations start to increase after a certain pulse width. The increase in variations is due to the onset of domain nucleation (Fig. 5(d) 6 and 4V at 40ns). Even though the pulse amplitude is constant, the evolution of underlying domain structures causes the onset of nucleation with increasing $t_{pulse}$. In the multi-domain scenario, the field lines from the +P domains terminate in -P domains around domain walls [33]. These fields, known as the stray fields, help reduce the depolarization field in the FE layer. However, as the +P regions grow due to domain growth at low pulse widths (Fig. 5(d) 4 and 6V at 3ns), and come close to each other, the stray fields accumulate in the intermediate -P region from both the +P domains. This accumulation of stray fields causes the total electric field in the region to exceed the nucleation critical threshold and initiates the nucleation of new +P domains. For the pulse amplitudes falling in the range of coercive voltage distribution across the samples (4V), domain nucleation occurs in a subset of samples, which exacerbates the increase in variations.

Further, certain applications such as multi-state memories, synapses require achieving a partially switched (intermediate) polarization state across samples. The plot of the mean of $P_{R+}$ across samples versus pulse width ($\mu(P_{R+})$ vs. $t_{pulse}$ in Fig. 5(c)) indicates the possibility of reaching this target polarization state via various appropriate combinations of pulse widths and amplitudes. By 'appropriate combinations,' we mean that one can use either a higher amplitude pulse for a shorter duration or a lower amplitude pulse for a longer duration, as long as pulse amplitude exceeds a certain threshold such that the target polarization state can be reached. However, it is important to note that the device-to-device variations associated with these pulse combinations can be quite different and understanding the dependence of the variations on the pulse combinations plays a crucial role in optimizing the devices.

Therefore, we consider one such representative state (labeled as A in Fig. 5(c)) and analyze the variations associated with different pulse combinations that can be used to reach this state A. Our results (Fig. 6) indicate that while the final mean polarization is similar, the resulting device-to-device variations vary across these combinations. In particular, we find that using lower amplitude pulses for longer durations results in lower variations across samples compared to using higher amplitude pulses for shorter durations.

This is because, when using a higher amplitude pulse for a shorter duration, the underlying variations in the switching time across the samples come into play. The fast-responding samples switch a significant amount of polarization (due to higher $V_{pulse}$) while slow ones may not completely respond to the applied pulse. This leads to larger device-to-device variations. On the other hand, when using lower amplitude pulses for a longer duration, the variations in the switching times do not affect significantly as slow samples get more time to respond to the applied pulse. This results in reducing the overall device-to-device variations.

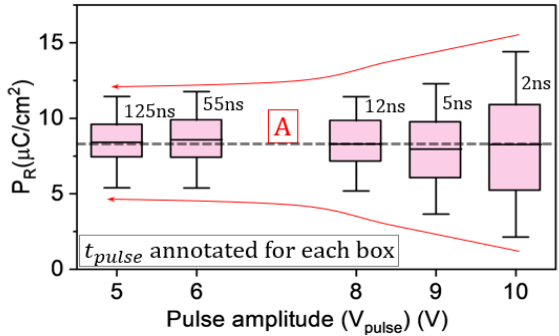

Fig. 6. Box plots of $P_R$ across the samples showing the variations and the pulse width required for different pulse amplitude to reach the target mean partially switched polarization state – A.

However, it should be noted that a decrease in pulse amplitude beyond a certain point yield diminishing returns i.e. we see little to no reduction in variations. Additionally, using lower amplitude pulses also comes with a trade-off as the pulse widths necessary to reach the target state increase drastically with decreasing pulse amplitudes. Therefore, to counter the device-to-device variations in $P_R$, reducing the pulse amplitude along with increasing the pulse width can be appealing but only up till a point beyond which the latency costs drastically increase with minimal improvement in variations.

## IV. CONCLUSION

We analyzed the effect of applied voltage on the polycrystalline-induced device-to-device variations in HZO-based MFIM stacks. Based on 3D multi-grain phase-field simulations, we showed a non-monotonic dependence of variations on the set voltage ($V_{SET}$), which is in good agreement with previous experiments. We correlated this trend in variations to the underlying polarization switching mechanisms and the collapse of domains. Further, we analyzed the impact of ferroelectric thickness scaling and discussed its utility in reducing the peak variations. Additionally, we analyzed the dependence of variations on voltage pulse parameters such as pulse width and amplitude. We demonstrated that for multi-state operation, when aiming for an intermediate or partially switched polarization state, using a lower amplitude pulse for longer duration results in reduced variations compared to a higher amplitude pulse for a shorter duration.